\documentclass[journal=jacsat,manuscript=article]{achemso}

\usepackage{chemformula} 
\usepackage{hyperref}
\usepackage[T1]{fontenc} 
\usepackage{subcaption}
\usepackage{titlesec}

\author{Prabeen Kumar Pattnayak}
\author{Aloke Kumar}
\author{Gaurav Tomar}
\email{*gtom@iisc.ac.in}
\affiliation[Indian Institute of Science]
{Department of Mechanical Engineering, Indian Institute of Science, Bengaluru, India}

\title[Diffusion of Star-Shaped Macromolecules]
  {Diffusion dynamics of star-shaped macromolecules in dilute solutions}

\abbreviations{IR,NMR,UV}

\begin{document}


\begin{abstract}

Polymer chains dissolved in a solvent take random conformations due to large internal degrees of freedom and are characterized geometrically by their average shape and size. The diffusive dynamics of such large macromolecules play an indispensable role in a plethora of engineering applications. The influence of the size of the polymer chain on its diffusion is well studied, whereas the same cannot be said for the shape of the polymer chain. In the present work, the influence of shape on the center-of-mass diffusion of the star-shaped chains in solution is investigated using Multi-particle Collision Dynamics. Star-shaped chains of varying degrees of functionality are modeled in a good solvent at infinite dilution. The radius of gyration($R_g$) of the star-shaped chains follows a functionality-independent scaling law with the chain length($N$), $R_g \sim N^{\nu}$, where $\nu \sim 0.627$. The shape of the polymer chains is calibrated by relative shape anisotropy. Highly anisotropic star-shaped polymer chains are found to have a faster rate of diffusion along the translational direction due to a slower rate of rotational diffusion when the radius of gyration of the polymer chains is maintained constant. 

\end{abstract}

\section{Introduction}

Polymeric fluids are a unique class of complex fluids that show a plethora of fascinating non-Newtonian behaviours\cite {bird1984fascinating}, which can be understood with the transport and rheological properties of the fluid. One key challenge of the complex fluid community is understanding how the macroscopic properties of the polymeric fluids arise from microscopic interactions of the macromolecules\cite{schroeder2018single}. The polymer chain dissolved in a solvent can take a multitude of conformations due to its large internal degrees of freedom. The average shape and size are used to characterize a polymer chain geometrically: the radius of gyration is widely used to describe the size and the eigenvalues of the gyration tensor for the shape calibration\cite{theodorou1985shape}. Advances in controlled polymerization have led to synthesizing complex polymer structures like star-shaped, comb-shaped, H-shaped, ring, and many more.\cite {fournier2007clicking}. The diffusive dynamics of such complex polymer chains are of fundamental interest in the biophysics community and are ubiquitous in numerous engineering applications. Star-shaped polymers are used for the controlled delivery of drugs in biomedical applications\cite{castano2015Polyglutamates} and as viscosity index modifiers in oil industries\cite{port1951viscosity}. Polyethylene glycol stars are used for protein delivery\cite{peppas1999poly}. Understanding macromolecular diffusion in a biological cell is important for its various functions, such as the movement of plasmids and transport of amino acids \cite{mika2011macromolecule} \cite{trevors2012perspective}. Hence, the influence of the shape and size of the polymer chain on its diffusive dynamics in solution is essential.

Most studies on the diffusion of complex polymer chains have been done by keeping the length of the polymer chain constant. Using fluorescence microscopy, Robertson et al.\cite{robertson2006diffusion} have shown a lower radius of gyration and a higher diffusion coefficient for circular DNA molecules than linear ones for the same chain length. Using Brownian dynamics simulations with hydrodynamic interaction, Kanaeda and Deguchi\cite{kanaeda2008diffusion} have reported higher diffusion coefficients for the ring polymers than for the linear polymers of the same chain length. Hegde et al.\cite{hegde2011conformation} have reported similar findings for the ring polymers in comparison to the linear chains for the same chain length by using three different simulation techniques. Singh et al.\cite{singh2014hydrodynamic} have shown, using Multi-particle Collision Dynamics(MPCD), that the center-of-mass diffusion coefficient of the star polymer chains decreases with an increase in their radius of gyration. Hence, when it comes to size, it is clear that the higher the size of the polymer chain, the lower the center-of-mass diffusion coefficient. However, it is difficult to comment on the influence of the shape of the polymer chain on its diffusion from the same polymer chain length study as both shape and size are distinct for different polymer chain architectures\cite{khabaz2014effect}. The diffusion study of complex polymer chains for the same size case is scanty. Hegde et al.\cite{hegde2011conformation} have reported a higher diffusion coefficient for the linear chains than the ring chains for the same radius of gyration using Molecular Dynamics, MPCD, and the Lattice Boltzmann method and also noted that size could not be the only factor that influences the diffusion of the chain. Therefore, the effect of the shape parameter on the center-of-mass diffusion of the polymer chains still remains an open question.

In this work, the effect of the shape parameter on the center-of-mass diffusion of the star-shaped polymer chains in solution is studied in the limit of infinite dilution using a mesoscopic coarse-grained simulation method, namely MPCD. For simulating the Brownian motion of the complex polymer chains in a solution, MPCD is frequently used as it incorporates both thermal fluctuation and long-range hydrodynamic interactions\cite{gompper2009multi}. At first, the shape and size of star-shaped polymer chains with different functionality are analyzed using the gyration tensor and compared with linear polymer chains at the same chain length. Subsequently, the translational diffusion of six different types of chains (one linear and five star-shaped chains) with the same radius of gyration is studied using the center-of-mass mean square displacement, followed by their rotational diffusion using the reorientation correlation function. Finally, the diffusion study is correlated to the shape characterization study in order to find the effect of the shape parameter on the center-of-mass diffusion of the star-shaped polymer chains in a solution.

\section{Numerical formulation}
 The coarse-grained bead-spring model represents the polymer chains dissolved in the solution. To replicate good solvent conditions, the excluded volume interactions between the monomer beads are modeled using the repulsive part of the 12-6 Lennard-Jones (LJ) potential, also known as Weeks-Chandler-Andersen potential\cite{weeks1971role} ($U_{WCA}$), defined as:
\begin{equation}
U_{WCA}(r) = 
    \begin{cases}
  4\varepsilon \left[ \left( \frac{\sigma_p}{r} \right)^{12} - \left( \frac{\sigma_p}{r} \right)^6 \right] + \varepsilon & $r $\leq 2^{1/6}\sigma_p$ $ \\
  0 & \text{otherwise}
\end{cases}
\end{equation}
 where $\sigma_p$ is the diameter of a monomer bead, $r$ is the distance between two beads and $\varepsilon = k_BT$ is the strength of interaction, $k_B$ is the Boltzmann’s constant and $T$ is temperature. The neighboring monomers of the polymer chain are connected with springs, the potential of which is given by Finitely Extensible Nonlinear Elastic (FENE)\cite{warner1972kinetic}, defined as:
\begin{equation}
    U_{FENE}(r) = 
\begin{cases}
  -\frac{1}{2} k r_0^2 \ln \left[ 1 - \left(\frac{r}{r_0}\right)^2 \right]   & $r $\leq r_0$ $ \\
  \infty & \text{otherwise}
\end{cases}
\end{equation}
where $k$ is the spring constant, and $r_0$ is the maximum length of the extension. The values of $\kappa$ and $r_0$ are 30 $\frac{k_BT}{\sigma_p^2}$ and $1.5 \sigma_p$ respectively, as recommended by Kremer and Grest\cite{kremer1990dynamics}. The bead spring model, FENE potential, and Kreme and Grest parameters are widely used in the coarse-grained modeling of the polymer chains\cite{gartner2019modeling}. The star-shaped polymer chains of varying degrees of functionality (number of arms) have been modeled by connecting different linear arms at their ends instead of connecting them to a single central monomer, ensuring equal flexibility of the arms for all the functionalities.

The solvent is modeled explicitly as an ensemble of non-interacting point particles of finite mass ($m$) using a mesoscopic coarse-grained simulation technique, MPCD\cite{malevanets1999mesoscopic}. MPCD consists of alternating streaming and collision steps. In the streaming step, the MPCD particles with velocity $\boldsymbol{v}_i$ undergo ballistic motion and their positions ($\boldsymbol{r}_i$) are updated as:
\begin{equation}
    \boldsymbol{r}_i( t + \delta t) = \boldsymbol{r}_i(t) + \delta t \boldsymbol{v}_i(t)
\end{equation}
In the collision step, the simulation box is divided into cubic cells of equal size($a$), and all the particles within a cell undergo stochastic collision. The collision of the MPCD particles is modeled using a momentum-conserving version of the Andersen Thermostat, also known as MPCD-AT\cite{allahyarov2002mesoscopic}, in which the particle velocities ($ \boldsymbol{v}_i$) are updated as:
\begin{equation}
    \boldsymbol{v}_i(t + \delta t) = \textbf{v}_{cm}(t) + \boldsymbol{v}_i^{ran} - \Delta \textbf{v}_{cm}^{ran}
\end{equation}
where $\textbf{v}_{cm}$ is the center-of-mass velocity of the collision cell, $\boldsymbol{v}_i^{ran}$ is a random velocity selected from a Maxwell-Boltzmann distribution, and $\Delta \textbf{v}_{cm}^{ran}$ is the change in center-of-mass velocity of the collision cell due to the addition of $\boldsymbol{v}_i^{ran}$. During the streaming interval of MPCD, the positions, and velocities of the monomer beads evolve by the velocity-Verlet algorithm\cite{allen1989computer} with a time step $\delta t_{MD}$. During the collision step, the monomers are considered MPCD particles and undergo stochastic collisions. The three components of $\boldsymbol{v}_i^{ran}$ are selected from a Gaussian distribution with variance $k_BT/m$ for the solvent particles and $k_BT/M$ for the monomer beads, where $M$ is the mass of a monomer. This way of considering the monomers just like other MPCD particles in the collision step for modeling solvent-monomer interaction is often used in recent studies\cite{hegde2011conformation}\cite{jiang2013accurate}\cite{nikoubashman2017equilibrium}\cite{chen2017effect}\cite{chen2018coupling} due to its advantage of avoiding spurious depletion forces\cite{padding2006hydrodynamic} which could lead to breakage of FENE bonds. Galilean invariance is ensured by randomly shifting the cells before each collision step by a vector with the three components randomly chosen from $[-a/2,a/2]$\cite{ihle2001stochastic}. All the simulations have been performed using the MPCD-AT routines in LAMMPS\cite{LAMMPS}(Chen et al.\cite{chen2018coupling} \cite{2017GitHub}).

\begin{figure}
  \includegraphics[scale=0.3]{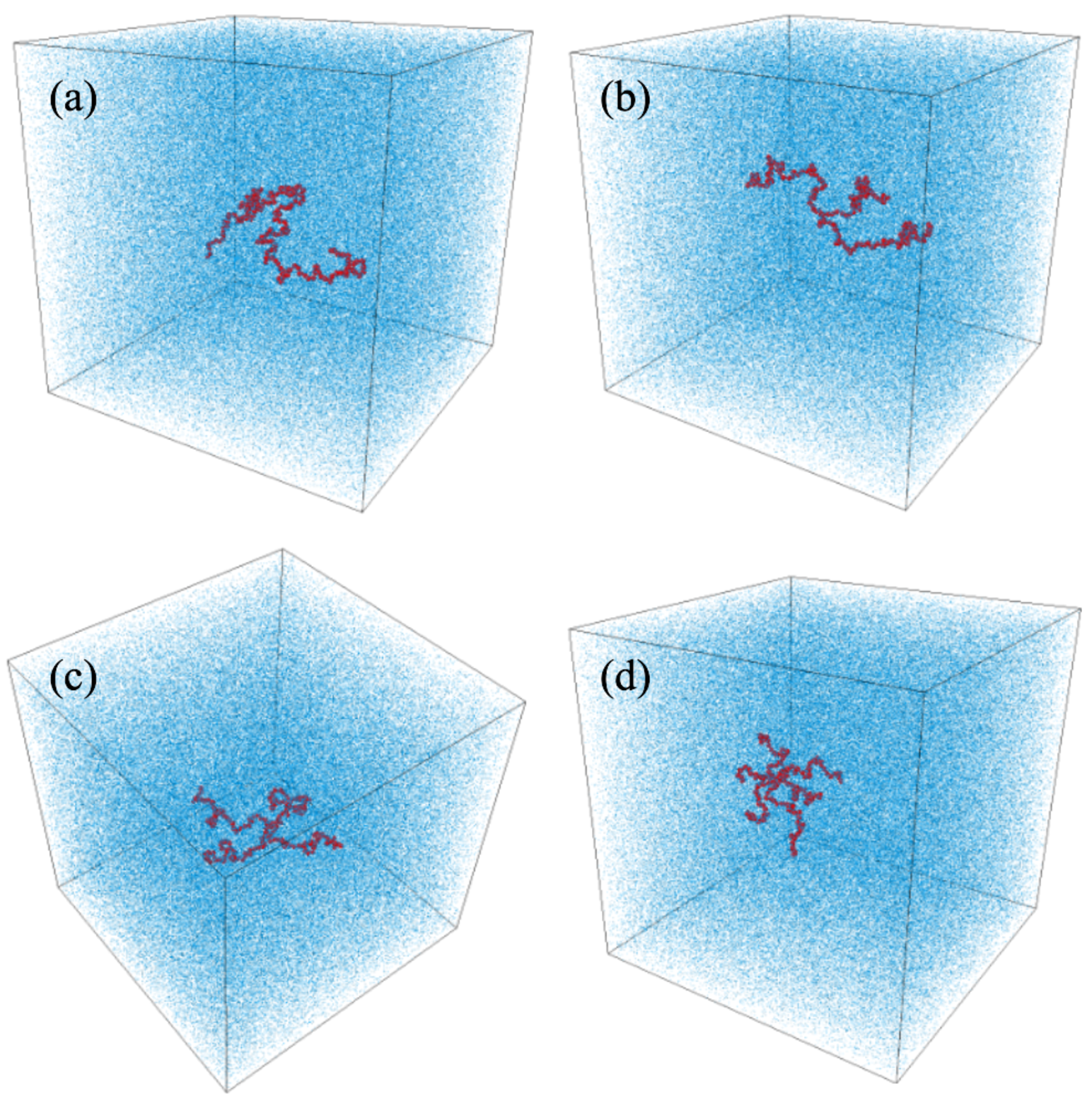}
  \caption{MPCD simulation snapshots: (a) Linear chain, (b) 3-armed star chain, (c) 4-armed star chain, and (d) 6-armed star chain. All four types of chains have the same chain length of 192$\sigma_p$, and the size of the cubic box is 64$\sigma_p$, where $\sigma_p$ is the diameter of a monomer bead. Monomer beads are represented by red color and solvent particles by blue. OVITO visualization tool is used. \cite{stukowski2009visualization}}
  \label{fig:snap}
\end{figure}

The size of the collision cells is taken to be the same as the size of the monomer beads, $a = \sigma_p$. The average density of the MPCD solvent equals 5$\frac{m}{\sigma_p^3}$. The mass of a monomer($M$) is taken as 5$m$ to achieve neutral buoyancy. The MD time step ($\delta t_{MD}$) equals $0.002\tau$. The MPCD collision time step ($\delta t$) is $0.09\tau$, where $\tau$ is the intrinsic unit of time equals $\sqrt{m\sigma_p^2/k_BT}$. The resulting viscosity and Schmidt number($Sc$) of the MPCD fluid are $4 \frac{k_BT}{\sigma_p^3}$ and 12, respectively. The size of the cubic simulation box is increased with polymer chain length to avoid the finite size effects following the previous studies.\cite{hegde2011conformation}\cite{chen2017effect} Box size($L$) equals $32\sigma_p$ for the set of chain lengths $\{ 24\sigma_p, 36 \sigma_p, 48\sigma_p \}$ , $48 \sigma_p$ for the set of chain lengths $\{ 60\sigma_p, 84 \sigma_p \}$, and $64\sigma_p$ for the set of chain lengths $\{ 108\sigma_p, 192 \sigma_p \}$. The equilibration simulation run is performed for $2\times10^6$ MD time steps. The results are time averaged over $5\times10^8$ MD time steps and ensemble-averaged over five system replicas, each with a unique set of random velocities at starting of the simulation and during the stochastic collision, both taken from  Maxwell-Boltzmann distribution. The measured parameters will be expressed in reduced units using the energy scale $k_BT$, length scale $\sigma_p$, and mass scale $m$. Periodic boundary conditions are implemented in all directions. The snapshots of the simulations are shown in Figure \ref{fig:snap}.

\begin{figure}[h]
\centering
\begin{subfigure}{.5\textwidth}
  \centering
  \includegraphics[scale=0.5]{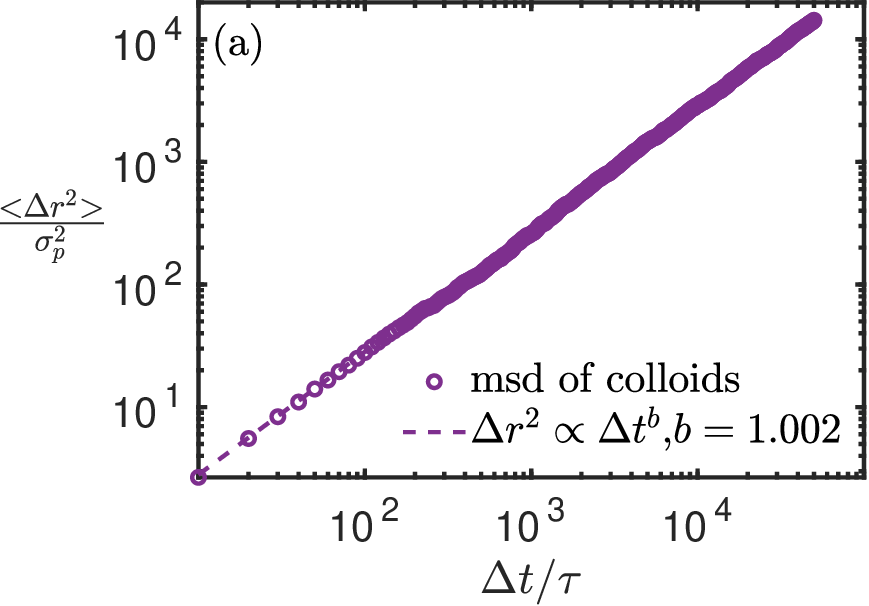}
  \label{fig:sub1}
\end{subfigure}%
\begin{subfigure}{.5\textwidth}
  \centering
  \includegraphics[scale=0.5]{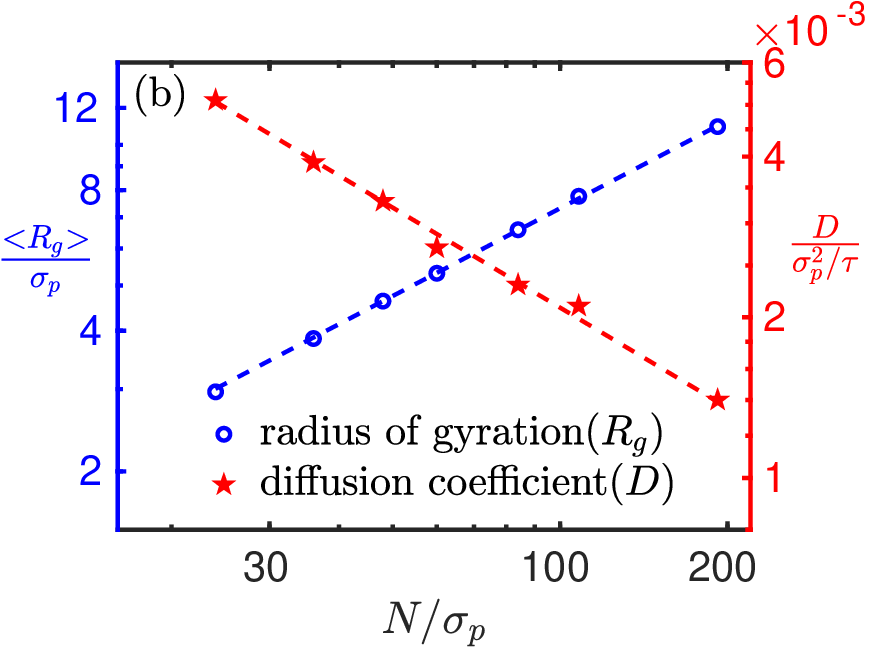}
  \label{fig:sub2}
\end{subfigure}
\caption{(a) Variation of MSD of colloidal particles in MPCD solvent vs. lag time. (b) Variation of radius of gyration and translational diffusion coefficient with chain length for linear chains with power-law fit}
\label{fig:test}
\end{figure}

To validate the MPCD routines, the Brownian motion of 250 colloidal particles of the same size as the monomer beads  are modeled in the MPCD solvent. The variation of their average mean square displacement (MSD) with lag time ($\Delta t$) is plotted in Figure \ref{fig:test}(a). Typically, power law describes the variation of MSD with lag time: MSD $\propto \Delta t^b$. The dynamics of the solutes can be diffusive ($b=1$), sub-diffusive ($b<1$), or super-diffusive ($b>1$). From Figure \ref{fig:test}(a), we note that we obtain b = 1 as expected for the colloids. Hence, the dynamics of the colloids are captured well by the simulation. Further, the radius of gyration($R_g$) vs. chain length ($N$) is plotted for the linear chain in Figure \ref{fig:test}(b). A power law behavior can be observed with a scaling exponent of 0.623 for the linear chain. The value of the power law exponent reported by Chen et al.\cite{chen2018coupling} using similar simulation parameters is 0.61. Linear chains have been studied widely, and their corresponding scaling exponent values for good solvent conditions are summarized in Table \ref{tbl:expn}. The exponent value calculated in the present work agrees well with earlier studies. In addition, the diffusion coefficient($D$) is calculated from the center-of-mass mean square displacement(MSD) vs. lag time plot for the linear chains of different lengths. The variation of $D$ with $N$ is shown in Figure \ref{fig:test}(b). It also follows a power law $D \sim N^{-\nu_d}$, where $\nu_d = 0.622$. The equality of $\nu$ with $\nu_d$ confirms the Zimm theory for the diffusion of a polymer chain with intra-chain hydrodynamic interactions, which predicts $D \sim \frac{1}{R_g}$. 

\begin{table}
  \caption{ A summary of the value of the scaling exponent $\nu$ for linear polymers from various literature sources}
  \label{tbl:expn}
  \begin{tabular}{lll}
    \hline
                    & Literature source & value of $\nu$   \\
    \hline
    Theory          & Flory theory\cite{rubinstein2003polymer}   &  0.6 \\
                    & Renormalization group theory\cite{witten1981spatial}  & 0.588  \\
                    & Perturbation calculations\cite{doi1986modern}  & 0.588  \\
    Experiments     & Synthetic polymers(neutron scattering)\cite{PhysRevLett.32.1170}   &  $0.59 \pm 0.06$ \\
                    & DNA molecules(fluorescence microscopy)\cite{smith1996dynamical}   & $0.611 \pm  0.016$  \\
                    & HPAM(intrinsic viscosity measurements)\cite{poling2015size}  & $ 0.62 $  \\
    Simulations     & Brownian dynamics\cite{hegde2011conformation}   &  $0.644 \pm 0.002$ \\
                    & MPCD\cite{hegde2011conformation}  & $0.62 \pm 0.02$  \\
                    & Lattice Boltzmann\cite{hegde2011conformation}  & $0.623 \pm 0.03$  \\
                    & MD\cite{khabaz2014effect}  & $0.597 \pm 0.03$  \\
                    & MPCD-AT\cite{chen2018coupling}  & $0.61$ \\
                    & Present Work &  $0.622$ \\
    \hline
  \end{tabular}
\end{table}

\begin{figure}
  \includegraphics[scale=0.18]{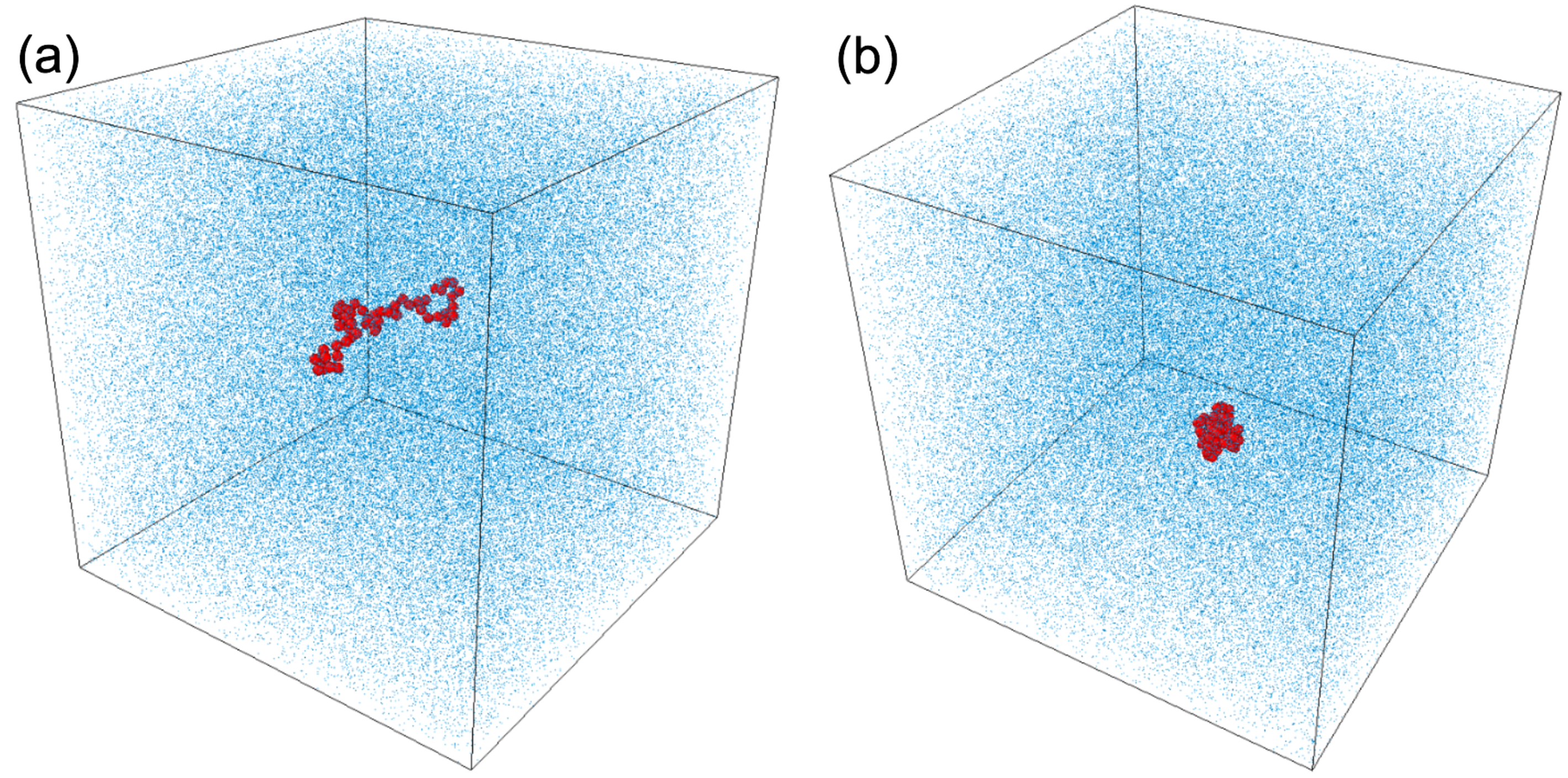}
  \caption{MPCD simulation snapshots: (a) Good solvent (b) Poor solvent. For both cases, the linear chain is of length $56\sigma_p$ in a cubic box of size 48$\sigma_p$, where $\sigma_p$ is the diameter of a monomer bead. OVITO visualization tool is used. \cite{stukowski2009visualization}}
  \label{fig:solvqual}
\end{figure}

The difference between a good solvent and a poor solvent can be demonstrated by introducing attraction using the standard 12-6 LJ potential with a cutoff of $2.5\sigma_p$ for the pairwise interaction between monomer beads instead of WCA potential as explained by Peng et al.\cite{peng2022reducing}. In a good solvent, the polymer chain forms a coil, whereas it becomes a globule in the case of a poor solvent. This coil-to-globule transition is shown in Figure \ref{fig:solvqual} and can be observed by measuring the radius of gyration, which is approximately $5\sigma_p$ in a good solvent for the chain length of $56\sigma_p$, contrary to $2\sigma_p$ for the poor solvent case. This reduction in the size of the polymer chain is also visible in the values of the resulting diffusion coefficients, which are $0.0061\sigma_p^2/\tau$ and $0.0032\sigma_p^2/\tau$ for poor and good solvents, respectively.

\section{Results and discussion}
\subsection{Shape and size of star-shaped chains}
The gyration tensor($\textbf{S}$) of a polymer chain is defined as the dyadic product of the position vector($\textbf{P}$) of a monomer bead in the center-of-mass reference frame with its transpose and averaged over all the monomers of the chain\cite{theodorou1985shape}.
\begin{equation}
\textbf{S} = \frac{1}{N}\textbf{P}\textbf{P}^T, \quad
    \textbf{P} = \begin{pmatrix}
    x_1^i - x_1^{cm} \\
    x_2^i - x_2^{cm} \\
    x_3^i - x_3^{cm} 
\end{pmatrix} 
\end{equation}
where ($x_1^{cm},x_2^{cm},x_3^{cm}$) represents the centre-of-mass of the polymer chain consisting of $N$ identical monomers ($x_1^i,x_2^i,x_3^i$) and is calculated as follows:
\begin{equation}
    x_1^{cm} = \frac{1}{N}\sum\limits_{i=1}^{N}x_1^i, \quad 
    x_2^{cm} = \frac{1}{N}\sum\limits_{i=1}^{N}x_2^i, \quad
    x_3^{cm} = \frac{1}{N}\sum\limits_{i=1}^{N}x_3^i
\end{equation}
The elements of $\textbf{S}$ can be written using indicial notation as:
\begin{equation}
    S_{pq} = \frac{1}{N} \sum\limits_{i=1}^{N} (x_p^i - x_p^{cm})(x_q^i - x_q^{cm})
\end{equation}
The polymer chain's shape and size can be easily measured by the eigenvalues of $\textbf{S}$, i.e., $\lambda_1, \lambda_2$, and $\lambda_3$. The radius of gyration($R_g$) represents the size of the polymer chain, the square of which is equal to the trace of the gyration tensor\cite{theodorou1985shape}. 
\begin{equation}
R_g^2 = Tr(\textbf{S}) = \lambda_1 + \lambda_2 + \lambda_3
\end{equation}

\begin{figure}[h]
\centering
\begin{subfigure}{.5\textwidth}
  \centering
  \includegraphics[scale=0.55]{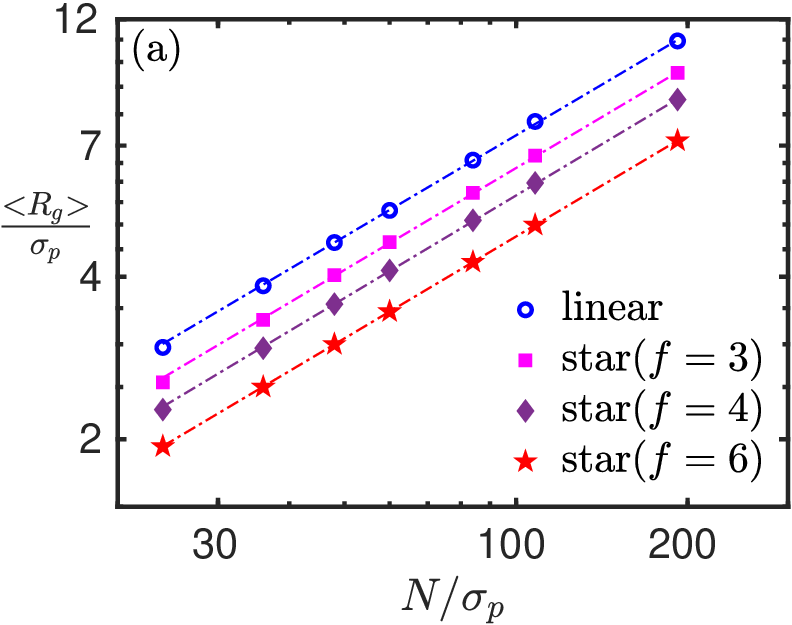}
\end{subfigure}%
\begin{subfigure}{.5\textwidth}
  \centering
  \includegraphics[scale=0.55]{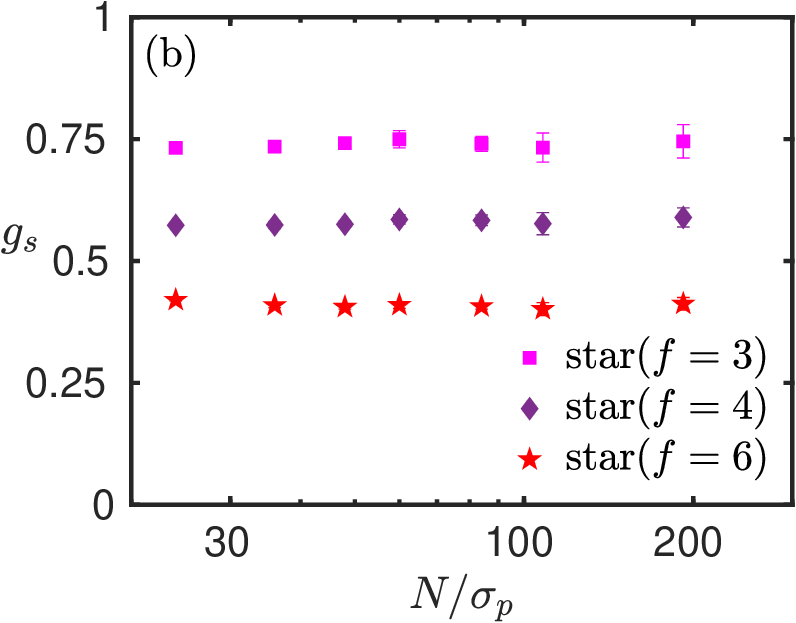}
\end{subfigure}
\caption{Variation of (a) radius of gyration and (b) geometric shrinking factor with chain length. The dotted lines represent power law fit, $R_g \propto N^{\nu}$ where the value of $\nu$ is 0.623, 0.627, 0.631, and 0.626 for the linear, 3-armed star, 4-armed star, and 6-armed star chain, respectively}
\label{fig:size}
\end{figure}

We evaluate $R_g$ for four different types of chain: linear, 3-armed star, 4-armed star, and 6-armed star using seven different chain lengths, and the results are summarized in Figure \ref{fig:size}(a). The radius of gyration follows a power law with polymer chain length, $R_g \sim N^{\nu}$. The power law's exponent($\nu$) represents the quality of the solvent. The value of $\nu$ calculated in the present simulations are 0.623, 0.627, 0.631, and 0.626 for the linear, 3-armed star, 4-armed star, and 6-armed star chains, respectively. The scaling law is found to be independent of the functionality of the star-shaped chains in which the average value of $\nu \sim 0.627$, indicating similar scaling behavior of the linear and star-shaped chains under good solvent conditions\cite{khabaz2014effect}\cite{daoud1982star}\cite{grest1987structure}. Table \ref{tbl:expn} summarizes the values of $\nu$ for linear chains reported by experiments, simulation, and theory. The calculated average value of $\nu$ is in good agreement with the existing results in the literature. For the same chain length, linear chains are bigger than the star-shaped chains, and among the star-shaped chains, size decreases with an increase in functionality, i.e. number of arms, as expected. Compared to the linear chain, this reduction in the size of the branched chains for the same polymer chain length is measured using the geometrical shrinking factor($g_s$) defined as the ratio of the mean squared radius of gyration of the branched chain to that of the linear chain, $g_s = \frac{\langle R_{g,b}^2 \rangle}{\langle R_{g,l}^2 \rangle}$\cite{khabaz2014effect}. The values of $g_s$ of star chains with different functionality and chain lengths are shown in Figure \ref{fig:size}(b). We note that $g_s$ doesn't vary much over the chain length for a particular chain type. The values of $g_s$ for star chain with $f=5$ reported by Khabaz and Khare\cite{khabaz2014effect} is approximately 0.5, which falls between 0.41(6-armed star) and 0.58(4-armed star) calculated in the present work and suggests linear variation in $g_s$ with $f$.

The shape of the polymer chains can be calibrated using asphericity($b$) and relative shape anisotropy($\kappa^2$) defined as,\cite{khabaz2014effect}\cite{theodorou1985shape}
\begin{equation}
b = \lambda_1 - \left( \frac{\lambda_2 + \lambda_3}{2} \right), \lambda_1 \geq \lambda_2 \geq \lambda_3
\end{equation}
\begin{equation} \label{eq:10}
    \kappa^2 = 1 - 3 \frac{\lambda_1 \lambda_2 + \lambda_2 \lambda_3 + \lambda_3 \lambda_1}{(\lambda_1 + \lambda_2 + \lambda_3)^2}
\end{equation}

\begin{figure}[h]
\centering
\begin{subfigure}{.5\textwidth}
  \centering
  \includegraphics[scale=0.55]{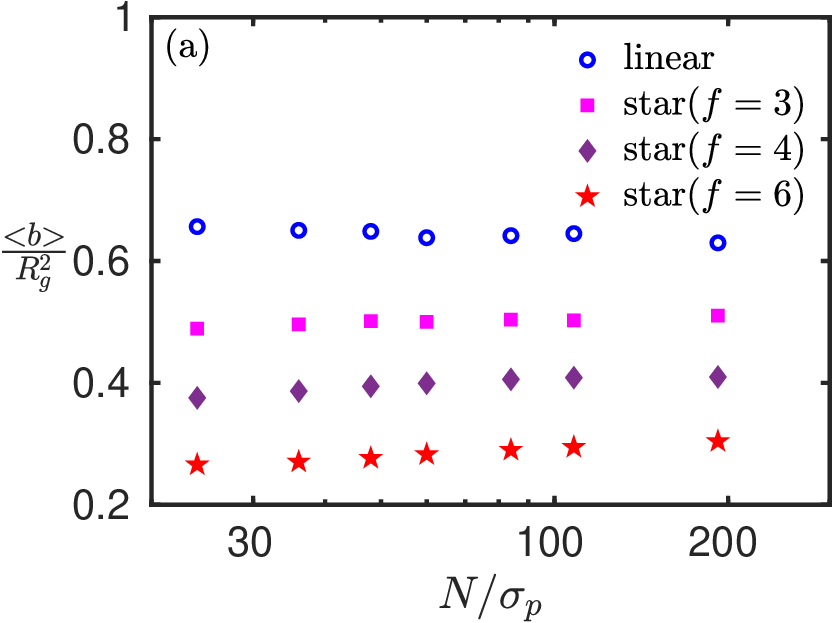}
  \label{fig:sub1}
\end{subfigure}%
\begin{subfigure}{.5\textwidth}
  \centering
  \includegraphics[scale=0.55]{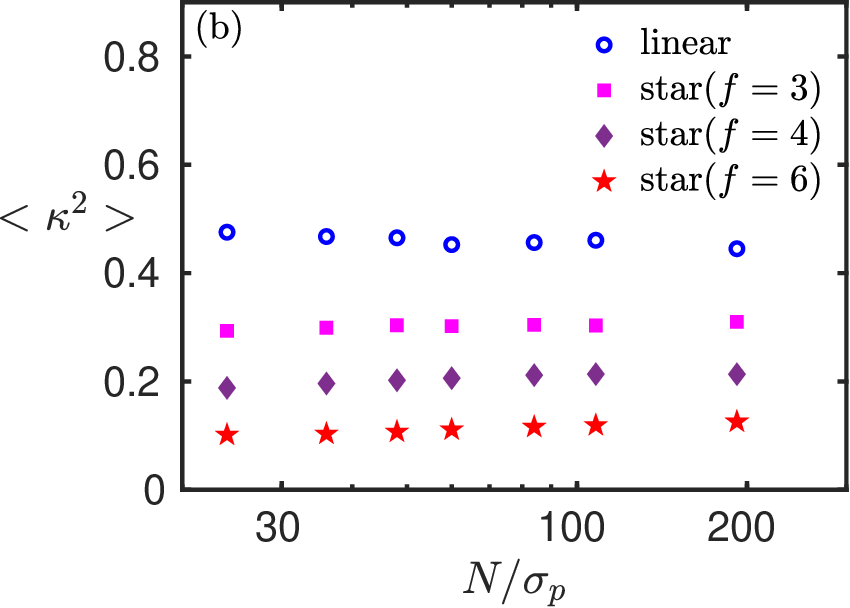}
  \label{fig:sub2}
\end{subfigure}
\caption{Variation of (a) normalized asphericity, $b/R_g^2$  and (b) relative shape anisotropy, $\kappa^2$ with chain length for linear and star-shaped polymer chains.}
\label{fig:shape}
\end{figure}

The variation of the two shape parameters with the polymer chain length is given in figure \ref{fig:shape}. The asphericity values are normalized by $R_g^2$ to make these independent of size.  For the individual architectures, the values of both shape parameters do not vary much over the chain lengths. The normalized asphericity can take any value between 0 and 1. It will be 0 for a spherical shape or any shape of the platonic solids and 1 for a rod-like structure. As expected, the linear chains have higher asphericity values than the star-shaped chains. Its value is $\sim$ 0.64 for the linear chain in the present work, which is close to 0.625(calculated using eigenvalues), reported by Koyama\cite{koyama1968excluded} and 0.66, reported by Theodorou and Suter\cite{theodorou1985shape}. Among the star-shaped chains, the normalized asphericity decreases with an increase in functionality. Khabaz and Khare\cite{khabaz2014effect} have reported $b/R_g^2$ $\sim$ 0.362 for the 5-armed star chain, which falls between 0.29(6-armed star) and 0.4(4-armed star) in the present work. The other shape parameter is the relative shape anisotropy($\kappa^2$), which also varies between 0 and 1. Its value is 1 for a rigid rod and 0 for a sphere and all platonic solids. It is also higher for linear chains than the star-shaped ones, as expected. In the present work, the value of $\kappa^2$ for the linear chain is $\sim$ 0.45, which  is nearly the same as 0.44 reported by Khabaz and Khare\cite{khabaz2014effect}. As anticipated, for star-shaped chains, $\kappa^2$ decreases with increasing functionality. In the present work, $\kappa^2$ $\sim$ 0.3(3-armed star), 0.21(4-armed star), and 0.12(6-armed star) which are slightly lower than 0.3454(3-armed star), 0.2463(4-armed star), and 0.1512(6-armed star), respectively, reported by Zifferer\cite{zifferer1999monte}. The value of $\kappa^2$ reported by Khabaz and Khare\cite{khabaz2014effect} for a 5-armed star chain is $\sim$ 0.16,  which falls between 0.12(6-armed star) and 0.21(4-armed star) calculated in the present work. To summarize, linear chains are less spherical and more anisotropic than the star-shaped chains. The higher the functionality among the star-shaped chains, the more spherical and less anisotropic the chain is. The variation of the shape parameters with functionality is plotted in Figure \ref{fig:corr}(a) and correlated with the diffusion coefficients in a later section.

\subsection{Translational diffusion of star-shaped chains}

\begin{figure}[h]
\centering
\begin{subfigure}{.5\textwidth}
  \centering
  \includegraphics[scale=0.55]{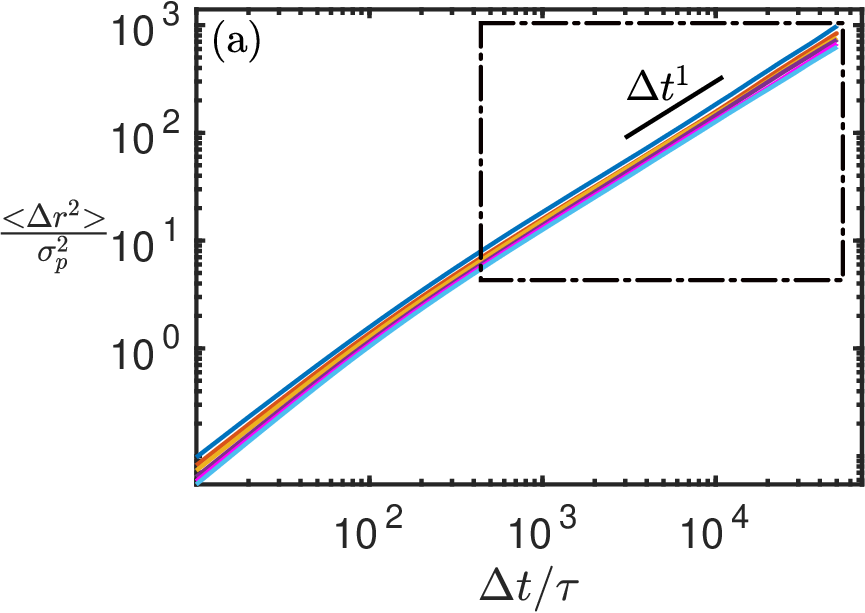}
  \label{fig:sub21}
\end{subfigure}%
\begin{subfigure}{.5\textwidth}
  \centering
  \includegraphics[scale=0.55]{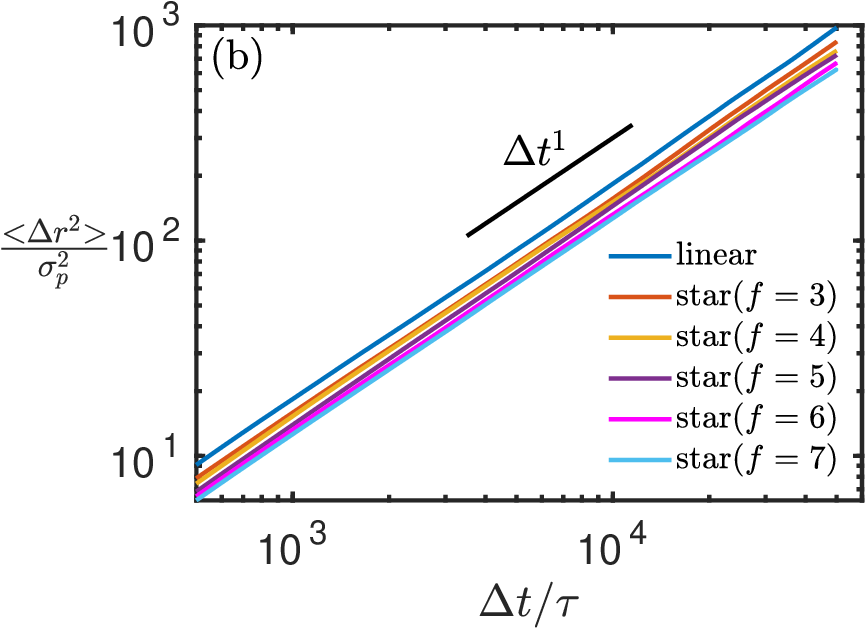}
  \label{fig:sub22}
\end{subfigure}
\caption{(a)Variation of center-of-mass mean square displacement, $\Delta r^2$ with lag time for different chains having approximately the same radius of gyration, $5\sigma_p$. (b) The diffusive regime (marked by the dotted rectangular region in (a)).}
\label{fig:diff}
\end{figure}

The diffusion rate of a polymer chain in a solution can be measured by the variation of center-of-mass mean square displacement(MSD) with lag time. One linear and five star-shaped chains are modeled to investigate the influence of the shape of the polymer chain on its diffusion. The effect of chain size is eliminated by selecting the polymer's chain length ($N$) such that the resulting radius of gyration is approximately $5\sigma_p$ for all six types. The simulation box size is $48\sigma_p$ for all six cases. The MSD($\Delta r^2$) vs. lag time($\Delta t$) plot is shown in Figure \ref{fig:diff}. At short times(less than $400 \tau $), MSD increases at a stronger rate than linearly with time due to the inertia of the chain. For longer times, the MSD reaches the linear diffusive regime, from which diffusion coefficients($D$) are calculated using the relation, $\Delta r^2 = 6D \Delta t$  and summarized in the second last column of Table \ref{tbl:samerg}.

\begin{table}
  \caption{Values of functionality($f$), arm length($N_a$), chain length($N$), radius of gyration($R_g$), normalized asphericity($b/R_g^2$), shape anisotropy($\kappa^2$), translational diffusion coefficient($D$) and rotational diffusion coefficient($D_R$) for star-shaped chains of different functionalities with same size ($R_g \sim 5\sigma_p$).}
  \label{tbl:samerg}
  \begin{tabular}{ccccccccc}
    \hline
    $f$ &$N_a/\sigma_p$& $N/\sigma_p$ & $<R_g/\sigma_p>$ & $<b/R_g^2>$ & $<\kappa^2>$ & $D/(\sigma_p^2/\tau)$ & $D_R \tau$ \\
    \hline
    2 & 28 & $56$    &  $5.08 \pm 0.52$   & $0.64 \pm 0.09$ & $0.46 \pm 0.12$ & 0.0032 & $1.4*10^{-4}$ \\
    3 & 23 & $69$    &  $5.04 \pm 0.2$    & $0.51 \pm 0.03$ & $0.31 \pm 0.03$ & 0.0028 & $2.2*10^{-4}$ \\
    4 & 20 & $80$    &  $5.02 \pm 0.15$   & $0.42 \pm 0.03$ & $0.23 \pm 0.03$ & 0.0026 & $3.2*10^{-4}$ \\
    5 & 19 & $95$    &  $5.00 \pm 0.08$   & $0.34 \pm 0.02$ & $0.16 \pm 0.02$ & 0.0025 & $3.6*10^{-4}$ \\
    6 & 18 & $108$   &  $4.98 \pm 0.07$   & $0.32 \pm 0.03$ & $0.13 \pm 0.02$ & 0.0022 & $4.2*10^{-4}$ \\
    7 & 17 & $119$   &  $4.95 \pm 0.08$   & $0.25 \pm 0.02$ & $0.09 \pm 0.02$ & 0.0021 & $4.8*10^{-4}$ \\
    \hline
  \end{tabular}
\end{table}

\begin{figure}
  \includegraphics[scale=0.55]{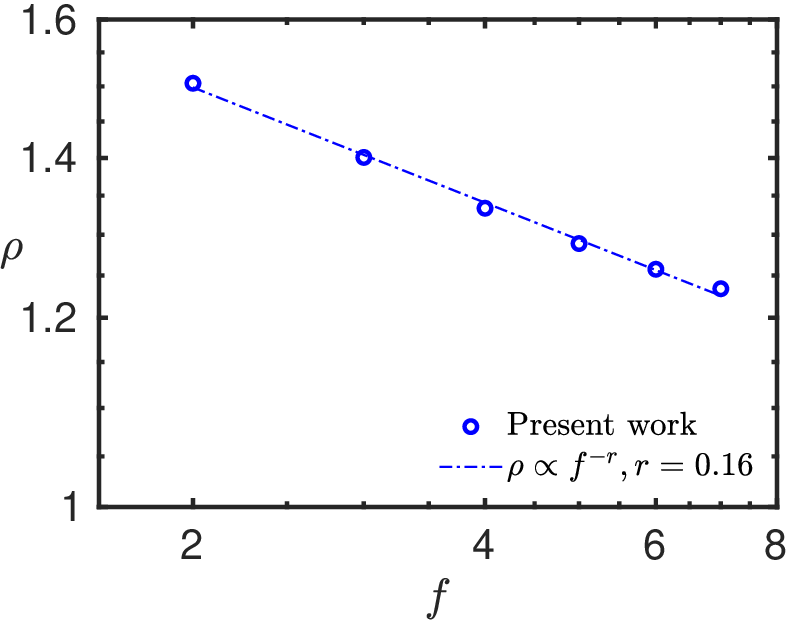}
  \caption{Variation of the ratio of the radius of gyration to hydrodynamic radius with the functionality of the star polymers.}
  \label{fig:rgrh}
\end{figure}

The linear chain can be considered a star chain with $f=2$ and has the highest value of diffusion coefficient. Among the star chains, the diffusion coefficient value decreases with increased functionality. Using the values of the diffusion coefficients, the hydrodynamic radius of the polymer chains can be calculated as\cite{burchard1980static},

\begin{equation}
    R_H = \frac{k_BT}{6 \pi \eta D}
\end{equation}
where $D$ is the translational diffusion coefficient of the polymer chain, and $\eta$ is the solvent viscosity. The ratio of the radius of gyration and hydrodynamic radius, $\rho = \frac{R_g}{R_H}$, is a size-independent quantity and represents the effect of the architecture of the polymer chain on its diffusion. The variation of $\rho$ with the functionality of the star polymers is plotted in Figure \ref{fig:rgrh}. We note that $\rho$ decreases with an increase in the functionality of the star chain, as reported by Huber et al. \cite{huber1984dynamic} and Singh et al. \cite{singh2014hydrodynamic}. Since all six types of chains are of the same size, this difference in the diffusion coefficient values can only be attributed to their shape parameters. In Figure \ref{fig:shape}, we have shown that the linear chain is more anisotropic and less spherical than the star-shaped chains, and among the star chains, $\kappa^2$ and $b/R_g^2$ decrease with increased functionality. Hence, the higher a star chain's relative shape anisotropy and normalized asphericity, the faster it diffuses along the translational direction. We investigate this further by computing the rotational diffusion of the polymer chains.

\subsection{Rotational diffusion of star-shaped chains}
\begin{figure}
  \includegraphics[scale=0.2]{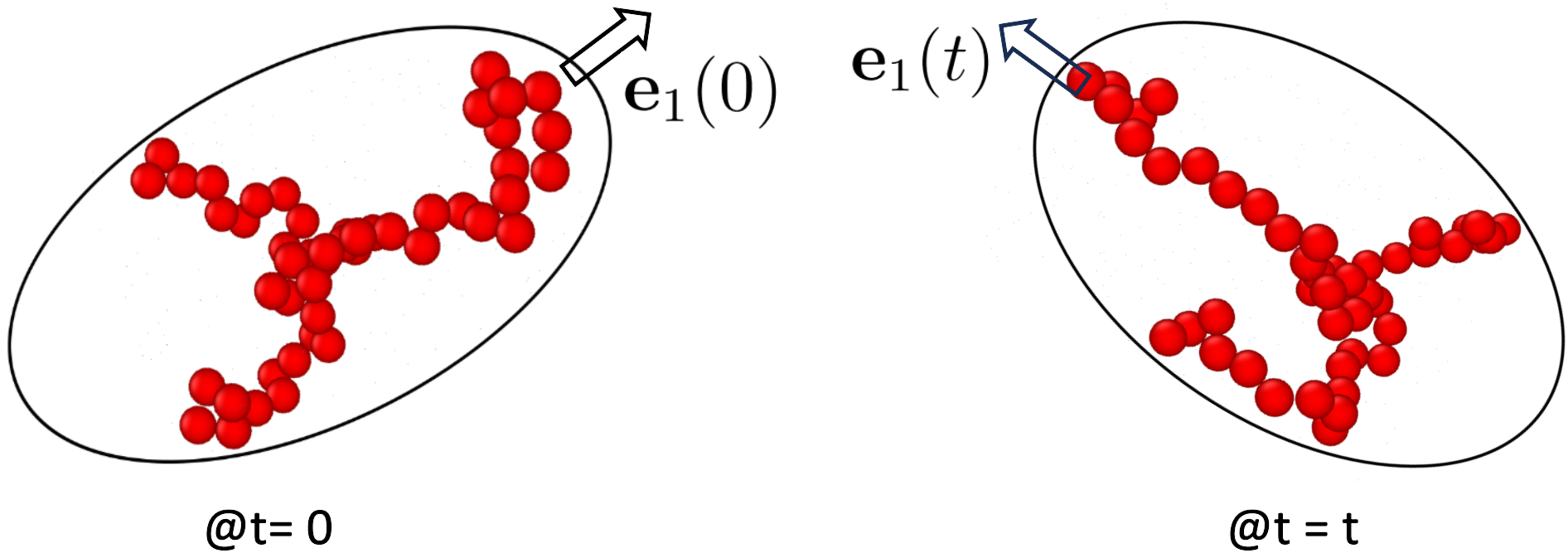}
  \caption{Schematic representation of the imaginary ellipsoidal shape surrounding the polymer chain. The two imaginary ellipsoids represent the orientation of the polymer chain at two different instants of time. (Reorientation of the polymer chain is equivalent to the rotation of the imaginary ellipsoid).}
  \label{fig:ellip}
\end{figure}

The polymer chain reorients itself continuously in the solution while diffusing along the translational direction. The gyration tensor has real eigenvalues and orthogonal eigenvectors as it is a symmetric tensor approximating the polymer chain as an ellipsoidal shape\cite{theodorou1985shape}. The reorientation of the polymer chain is equivalent to the rotation of the imaginary ellipsoid, as explained using a schematic representation in Figure \ref{fig:ellip}. Any vector rigidly attached to the polymer chain can be used for measuring the rate of reorientation. In this work, the eigenvector($\textbf{e}_1$) corresponding to the largest eigenvalue($\lambda_1$) of the gyration tensor is selected for measuring the rate of reorientation of the corresponding polymer chain. The relevant reorientational correlation function of the polymer chain can be defined as,
\begin{equation}
    C(t) = \langle P_2(\textbf{e}_1(0).\textbf{e}_1(t)) \rangle
\end{equation}
where $P_2(x)=(3x^2 - 1)/2$, is the second-order Legendre polynomial, and the angle bracket represents the time and ensemble average over five system replicas. For any isotropically reorienting polymer chain, following Wong et al.\cite{wong2009influence}, the reorientational correlation function can be approximated as,
\begin{equation}
    C(t) = e^{-6D_Rt}
\end{equation}
where $D_R$ is the rotational diffusion coefficient of the polymer chain.

\begin{figure}
  \includegraphics[scale=0.55]{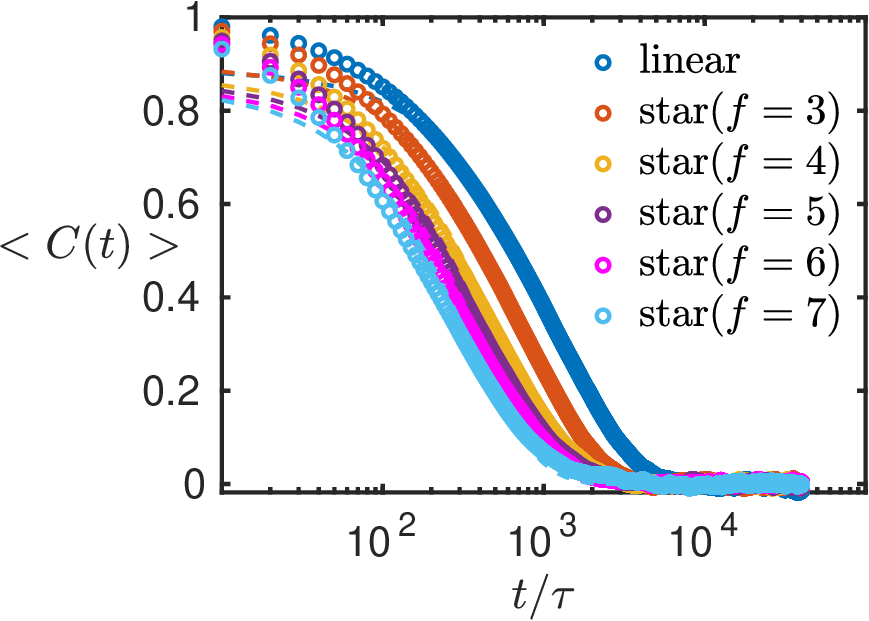}
  \caption{Variation in reorientational correlation function with time for different types of chains with approximately the same radius of gyration, $5\sigma_p$, The dotted curves are exponential fits.}
  \label{fig:rot}
\end{figure}

The variation of the reorientational correlation function with time is plotted in Figure \ref{fig:rot}. We note that $C(t)$ decays faster for the star-shaped chains than the linear chain. For the star-shaped chains, the higher the functionality, the faster the decay of $C(t)$. The rotational diffusion coefficients are calculated from the exponential fit using the least square method and are summarized in the last column of Table \ref{tbl:samerg}. The corresponding coefficient of determination($R^2$) is more than 0.99 for all the cases. The faster the decay of the reorientational correlation function, the higher the value of $D_R$. The linear chain has the lowest value of $D_R$, and among the star chains, $D_R$ increases with increased functionality. As discussed earlier for translational diffusion, this difference in the values of the rotational diffusion coefficient is because of the shape parameters, as all the six types of polymer chains considered here are of the same size. In terms of shape parameters, star polymer chains with lower values of relative shape anisotropy and normalized asphericity have a higher rate of rotational diffusion. The lower values of $\kappa^2$ and $b/R_g^2$ represent higher symmetry of monomer distribution with respect to the coordinate axes. It is intuitive that a star-shaped chain reorients faster when the distribution of its monomers is symmetrical with respect to the coordinate axes. It is to be noted that the variation in the rotational diffusion coefficient with functionality and shape parameters is opposite to that of the translational diffusion coefficient.

\subsection{Correlation of diffusion and shape parameters of star-shaped chains}

\begin{figure}[h]
\centering
\begin{subfigure}{.5\textwidth}
  \centering
  \includegraphics[scale=0.54]{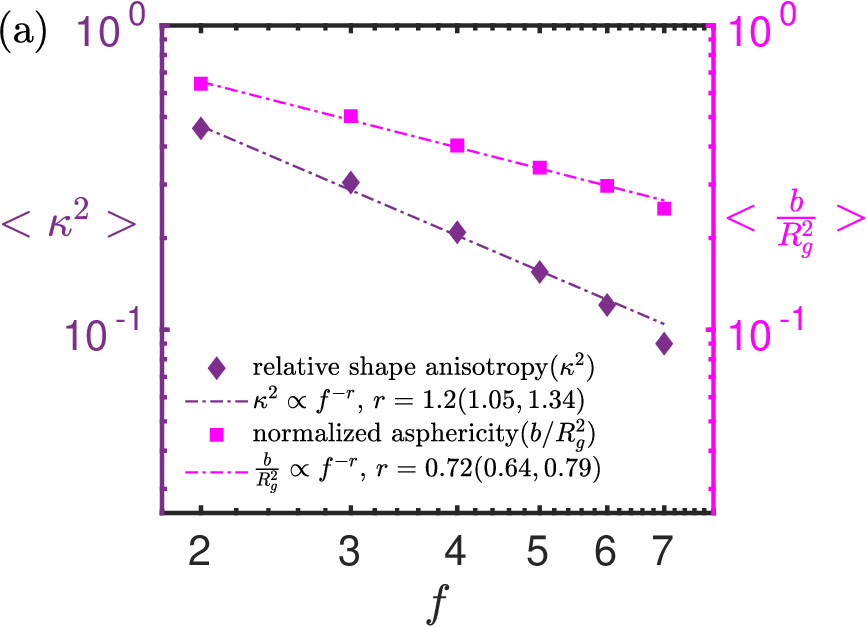}
  \label{fig:sub21}
\end{subfigure}%
\begin{subfigure}{.5\textwidth}
  \centering
  \includegraphics[scale=0.54]{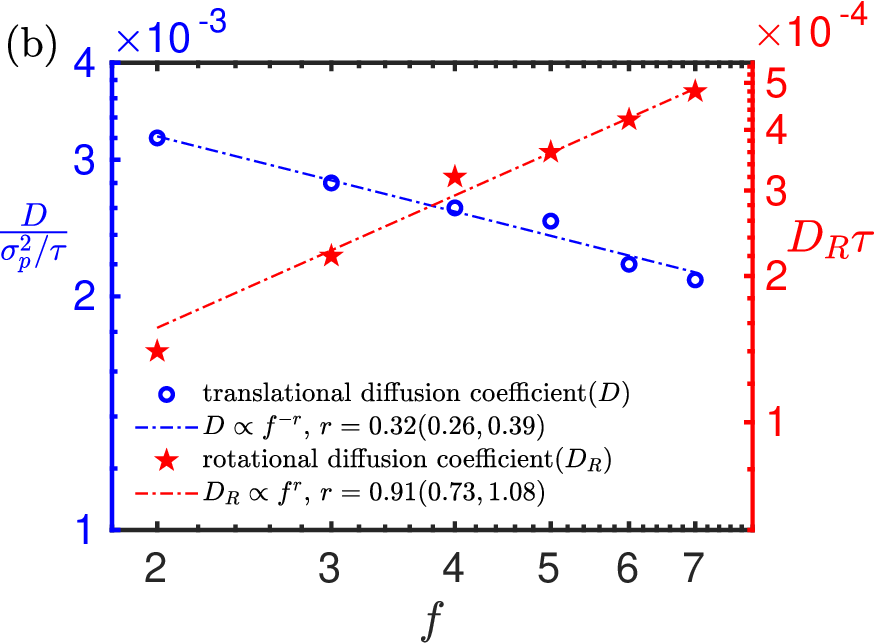}
  \label{fig:sub22}
\end{subfigure}
\caption{Variation of (a) shape parameters and (b) two types of diffusion coefficients with the functionality of star-shaped chains, where the values within the parentheses are the confidence intervals for the exponent, $r$ calculated from power law fits. The linear chain is considered a star-shaped chain with $f=2$.}
\label{fig:corr}
\end{figure}

The variation of the shape parameters and the two types of diffusion coefficients with the functionality of the chains are plotted in Figure \ref{fig:corr}(a) and Figure \ref{fig:corr}(b), respectively, where the linear chain is considered a star chain with $f = 2$. Out of the two shape parameters, relative shape anisotropy($\kappa^2$) can be expressed in terms of the invariants of the gyration tensor and is the overall measure of shape anisotropy\cite{theodorou1985shape}. By making two-to-one correspondence between the two types of diffusion coefficients in Figure \ref{fig:corr}(b) and the relative shape anisotropy in Figure \ref{fig:corr}(a), it can be stated that, for star-shaped chains with higher $\kappa^2$ values, the value of $D$ is higher, and the value of $D_R$ is lower. The origin of a polymer chain's translational and rotational diffusive motion is the collision with the surrounding solvent particles. The radius of gyration can be interpreted as the radius of the imaginary sphere surrounding the polymer chain in the solution. Maintaining the same $R_g$ for all six types of chains leads to the same size of the corresponding imaginary sphere. Therefore, all six types of chains interact with an approximately equal number of solvent particles on average. Highly spherical and isotropic star-shaped polymer chains utilize more energy in rotational diffusion, which results in less energy for diffusing along the translational direction. The opposite is the case for highly anisotropic star-shaped chains. Hence, the higher the relative shape anisotropy value of a star-shaped chain, the slower the rotational diffusion rate and the faster the rate of translational diffusion, as shown in Figure \ref{fig:corr}(b). 

\begin{figure}
  \includegraphics[scale=0.6]{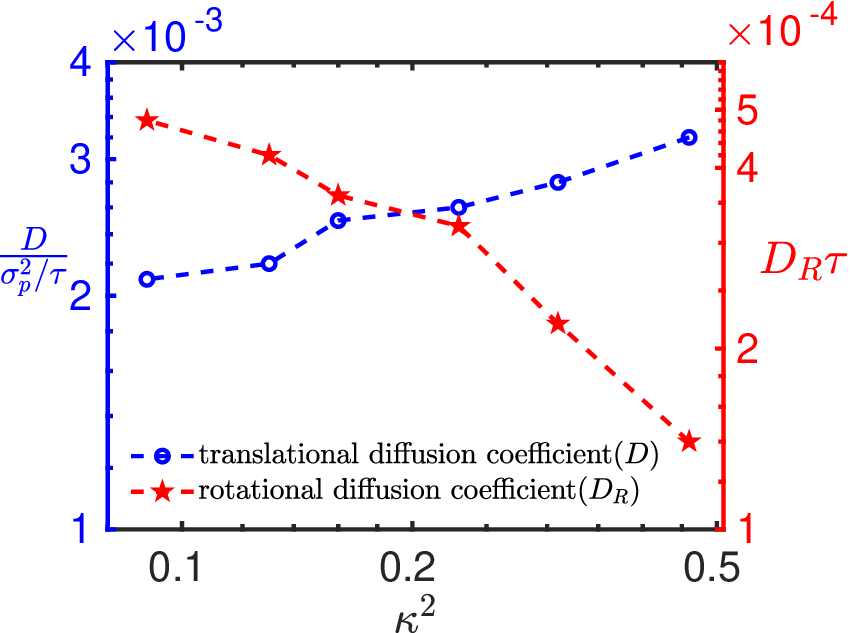}
  \caption{Variation of the translational diffusion coefficient and rotational diffusion coefficient with relative shape anisotropy of the star chains with approximately the same radius of gyration, $5\sigma_p$.}
  \label{fig:last}
\end{figure}

The variation of the translational diffusion coefficient and the rotational diffusion coefficient with relative shape anisotropy of the star-shaped chains having the same value of $R_g$ is shown in Figure \ref{fig:last}. The higher values of $\kappa^2$ lead to a lower value of the rotational diffusion coefficient and a higher value of the translational diffusion coefficient. From these results, we conclude that a star polymer chain with a higher value of relative shape anisotropy will have a slower rate of rotational diffusion and diffuse faster in the translational direction. Even though this is demonstrated using star-shaped chains, the argument can be extended to other polymer configurations as well. Hegde et al. have reported that the linear chains have higher translational diffusion coefficients than the ring chains when both have the same radius of gyration\cite{hegde2011conformation}. From the definition of relative shape anisotropy(equation \ref{eq:10}), it is intuitive that the linear polymer chain will have a higher value of $\kappa^2$ than the ring polymer chain. Hence, our argument also holds for the ring vs. linear case. Nevertheless, to verify this argument for generic polymer configurations, the study of other polymer chain architectures is essential.

\section{Conclusion}
In this work, the Brownian diffusion of the linear and star-shaped polymer chains of different functionalities is simulated using MPCD. It is shown that the radius of gyration of the star-shaped polymer chains follows a functionality-independent scaling law with chain length, in which the scaling exponent $\nu \sim 0.627$. The linear chain is shown to be more anisotropic than the star-shaped chains, and for star-shaped chains, the value of relative shape anisotropy decreases with an increase in functionality. For the same radius of gyration, the linear chain diffuses at a faster rate along the translational direction and has a slower rate of rotational diffusion than the star-shaped chains. Among star-shaped chains with the same radius of gyration, higher functionality leads to a higher value of rotational diffusion coefficient and a slower rate of diffusion along the translational direction. In terms of the shape parameter, we conclude that the star-shaped chains with higher values of relative shape anisotropy have a slower rate of rotational diffusion and therefore diffuse at a faster rate along the translational direction. Hence, shape anisotropy leads to faster center-of-mass diffusion of star-shaped polymer chains in a solution.


\begin{acknowledgement}
G.T. acknowledges partial support from the Department of Science and Technology National Supercomputing Mission HPC system in the Supercomputing Education and Research Center-Indian Institute of Science.  A.K. acknowledges partial support from SERB CRG/2022/005381. P.K.P. acknowledges partial support from the Ministry of Education, Government of India.

\end{acknowledgement}




\bibliography{achemso-demo}

\end{document}